\documentclass[epj,twocolumn]{webofc}
\usepackage{color}
\usepackage{soul}
\usepackage{ulem}
\usepackage[varg]{txfonts}  \woctitle{XLIV International Symposium on Multiparticle Dynamics}
\begin{document}
\title{$p-air$ production cross-section and  uncorrelated mini-jets processes in pp-scattering}

\author{Daniel A.  Fagundes\inst{1}\fnsep\thanks{\email{dfagundes@ift.unesp.br}} \and
       Agnes Grau\inst{2}\fnsep\thanks{\email{igrau@ugr.es}} \and
        Giulia Pancheri \inst{3}\fnsep\thanks{\email{giulia.pancheri@lnf.infn.it}}
             \and
        Yogendra N. Srivastava\inst{4}\fnsep\thanks{\email{yogendra.srivastava@gmail.com}}
             \and
        Olga Shekhovtsova\inst{5}\fnsep\thanks{\email{olga.shekhovtsova@ifj.edu.pl}}
}

\institute{Instituto de F\'isica Te\'orica, UNESP, Rua Dr. Bento T. Ferraz, 271, Bloco II, 01140-070, S\~ao Paulo - SP, Brazil 
\and
           Departamento de F\'{\i}sica Te\'orica y del Cosmos, Universidad de Granada, 18071 Granada, Spain
\and
           INFN Frascati National Laboratories, Via E. Fermi 40, Frascati I00044, Italy
\and Physics Department, University of Perugia, Perugia, Italy
\and Kharkov Institute of Physics and Technology
61108, Akademicheskaya,1, Kharkov, Ukraine\\
Institute of Nuclear Physics PAN ul. Radzikowskiego 152 31-342 Krakow, Poland
}

\abstract
{For the $p-air $ production cross-section, 
we use  a Glauber formalism which 
inputs the  $pp$ inelastic cross-section   from  a mini-jet model embedded 
in a single-channel eikonal expression, that provides the needed  
contribution of uncorrelated processes. It is then shown that current LO parton density functions for the 
$pp$ mini-jet cross-sections, with a rise tempered by  acollinearity induced by soft gluon re-summation,
are well suited to reproduce   recent cosmic  ray results.  By comparing results for GRV, 
MRST72 and MSTW parametrizations, we estimate  the uncertainty related to the low-x behavior of these densities.
}

\maketitle


%
\section{Introduction}
\label{intro}
In this contribution we address the problem of how to  relate accelerator  data for 
$proton- proton $ scattering to $p-air$ production 
cross-section measurements from cosmic rays.
This is a very old  question \cite{Glauber59,Gribov69} and very ingenious ways to do so  have been 
developed through the years \cite{Block:2006hy}. The  issue is often obfuscated 
by the need to estimate the contribution from elastic and diffractive processes, 
both in $p -air$, but mostly in $pp$ collisions. 

The question  of whether it is $\sigma^{pp}_{total}$ or $\sigma^{pp}_{inel }$ which is input 
to the Glauber formalism was discussed in 
the context of heavy ion collisions  in \cite{Kopeliovich2003} and  in
high energy cosmic rays in
 \cite{Anchordoqui:2004xb}.
Presently, most current analyses define a $\sigma^{p-air}_{prod }$ through the inelastic cross-section, and a 
$\sigma^{p-air}_{inel }$ through the total $pp$ cross-section.  In
 either case, elastic and quasi-elastic contributions need to be subtracted 
 and a degree of uncertainty can arise from their parametrization.
The definition of inelastic cross-section is also affected by uncertainties, both
 theoretically and experimentally, as seen in LHC experiments with different 
 cuts in the forward region \cite{Antchev:2013haa}.

Here we shall show that the total $p-air$ production cross-section can be
obtained in a very direct way through the inelastic $pp$ cross-section resulting
from single-channel eikonal models. This formalism for the inelastic cross-section
provides a description of non-correlated inelastic processes \cite{Achilli:2011sw}, and thus avoids
the problem of how to model  diffraction and    elastic cross-section. The
description of the latter, including the elastic differential cross-section, is still
not resolved, and is obtained through various parametrizations. 
 A recent suggestion by the Telaviv group \cite{Gotsman:2013nya} has made  efforts in this direction. 
 Here we shall follow a different path.

It is important to stress that in the case of cosmic rays, first and foremost
one needs an eikonal function which gives a description of the total $pp$ cross-
section, through a good understanding of the underlying physics. In this paper
we describe $proton -air$ production cross-section up to the recent AUGER
measurement \cite{Collaboration:2012wt}, using the inelastic $pp$ cross-section obtained from a QCD
mini-jet model with soft gluon re-summation \cite{Godbole:2004kx,Grau:1999em}.

We have long advocated QCD mini-jets as the driving mechanism for the rise of all 
total cross-sections \cite{Pancheri:1986qg} and have proposed a 
mechanism  based on infrared gluon resummation to tame the excessive rise with energy of the 
mini-jet cross-sections \cite{Corsetti:1996wg}. 
While the mini-jet cross-sections are seen to rise like a power law in energy, the proposed soft guon resummation ansatz  introduces a cut-off at large impact parameter values and brings the total cross-section close to a saturation of the Froissart bound.
Thus, the emphasis of the present work is two fold. 
First to provide a good phenomenological description of cosmic $p-air$ production cross-sections 
through a successful well accepted formalism, such as in the Glauber theory \cite{Glauber59}. 
The second is to reconfirm that the rise of all total, elastic and inelastic cross-sections of
protons on protons, or protons on nuclei and other hadrons, have the same origin: a rising contribution
from the increasing number (with energy) of low-x gluons excited in the collision  \cite{Gaisser:1984pg}. 

Since the  '80s, many models have used mini-jets in total cross-section physics 
\cite{Durand:1988ax,Durand:1988cr,Block:2000gy} and more recently in  \cite{Giannini:2013jla}. 
In  most cases, the parton density functions [PDFs] are chosen  or parametrized {\it ad hoc}.   
However,  we believe that mini-jets can give interesting information only
 if  used in connection with  current LO parton densities, such as  available through 
 updated PDF libraries. As in any perturbative QCD calculation, this LO effect needs then 
 to be complemented by  other QCD effects, such as that  of very soft gluons 
arising from the QCD confinement potential \cite{Corsetti:1996wg}.

In this contribution, the focus is on reproducing the very high energy cosmic ray phenomena, as measured by the AUGER collaboration and beyond. At such energies, the proton can penetrate the nucleus and interact independently with each nucleon, in addition QCD effects, obfuscated at lower energies by screening effects,  are now important.
We shall use the 
simplest version of the
Glauber model \cite{Glauber:1970jm}, 
with the following  basic hypothesis when the target is a nucleus: i) for low transverse momentum collisions 
$p_t \lesssim (1\div2)\ GeV$, the incoming proton does not penetrate the air nucleus and basically scatters off the 
surface, whereas, as the transverse momentum increases, the proton penetrates the nucleus of atomic number 
$A$ and scatters off all the protons in the volume occupied by the nucleus. Thus the nuclear density seen by 
the incoming protons will only be proportional to $A^{2/3}$ for the soft collisions, and  to $A$  for the hard 
part.
(ii) For interactions with
transverse momenta $p_t\gtrsim (1\div2)\ GeV$, we shall employ QCD effects in the form of mini-jets and soft gluon emission 
as in the model developed in
\cite{Godbole:2004kx,Grau:1999em,Corsetti:1996wg}.

Neglecting momentarily the above surface/volume effect, we begin 
with the usual Glauber expression for the production cross-section in the impact parameter representation,
as given by
\begin{equation}
\sigma_{prod}^{p-air}(E_{lab})=\int d^2\textbf{b} [1-e^{-n_{p-air}(b,s)}]
\end{equation}
with
\begin{equation}
n_{p-air}(b,s)=T_N(\textbf{b})\sigma_{inel}^{pp}(s)
\end{equation}
wherein $T_N(\textbf{b})$ is the nuclear density, for which we start by choosing a standard gaussian distribution, 
\begin{eqnarray}
T_{N}(b) = \frac{A}{\pi R_{N}^{2}}\textrm{e}^{-b^{2}/ R_{N}^{2}}\label{eq:nucl_prof}, 
\end{eqnarray}
properly normalized to
\begin{eqnarray}
\int d^{2} \textbf{b} T_{N}(b) = A.
\end{eqnarray}
The parameters used in the profile (\ref{eq:nucl_prof}), namely the average mass number of 
an ``air" nucleus, $A$, and the nuclear radius, $R_{N}$, are the following:
\begin{eqnarray}
A = 14.5 ,\quad  R_{N} = (1.1 fermi) A^{1/3}.
\end{eqnarray}
The inelastic $pp$ cross-section,  $\sigma_{inel}^{pp}$,  is obtained from $pp$ scattering, with 
\begin{eqnarray}
\sigma_{inel}^{pp}=\int d^2\textbf{b}[1-e^{-2\chi_I(b,s)}] \label{eq:siginel}\\
\sigma_{tot}^{pp}=2\int d^2 \textbf{b}[1- \Re e (e^{i\chi(b,s)})]\label{eq:sigtot}
\end{eqnarray}
where $\chi_I(b,s)=\Im m\chi(b,s)$ is the imaginary part of the eikonal function that 
defines the elastic amplitude. At high energy, it is a good approximation  to neglect a 
possible real part of the eikonal function in Eq.~(\ref{eq:sigtot}) and write
\begin{equation}
\sigma_{tot}^{pp}=2\int d^2 \textbf{b}[1-e^{-\chi_I(b,s)}]\label{eq:sigtotim}\\
\end{equation}
This formalism gives both the total and the inelastic  non-correlated cross-section, 
once the quantity $\chi_I(b,s)$ is known. The latter is an important point in the 
discussion of $p-air$ processes. 
The single-channel eikonal formalism for the inelastic cross-section
given by Eq.~(\ref{eq:siginel}) includes only non-correlated, Poisson distributed independent 
collisions. This can be seen easily by comparing this equation with a sum over all 
independent Poisson like distributions, as discussed in \cite{Achilli:2011sw}. 
Thus the above single-channel eikonal has the virtue of identifying all non-correlated processes, 
which we argue (and later verify phenomenologically) are all the non-diffractive processes 
contributing to the  $p-air$ production 
cross-section. We notice here that  this property  of the single-channel eikonal is a hindrance 
when one wants to separate the purely elastic from the diffractive part, but it is exactly what one 
needs for p-air shower initiated measurements. We shall return to this point again later. 
 
 In the following, we shall first consider $pp$ scattering and give a brief summary of the 
 physics content of our model and determine the parameters which give an 
 optimal description of $pp$ data up to LHC.
 We  shall then use  the single-channel eikonal to  calculate the inelastic non-diffractive $pp$ 
 cross-section and obtain the $p-air$ production cross-section to compare with data.
\section{Proton-proton total and inelastic non-diffractive cross-section}
The eikonal function of the mini-jet model of \cite{Godbole:2004kx,Grau:1999em} 
is given by
\begin{eqnarray}
2\chi_I(b,s)=n_{soft}^{pp}(b,s)+n_{jet}^{pp}(b,s)\nonumber\\
=A_{FF}(b)\sigma_{soft}^{pp}(s)+A_{BN}^{pp}(p;b,s)\sigma_{jet}(PDF,p_{tmin};s)
\end{eqnarray}
where $A_{FF}(b)$, the impact parameter distribution in the non perturbative term, is obtained through a  
convolution of two proton form factors, whereas for the perturbative term, the distribution $A_{BN}^{pp}(p;b,s)$,
multiplying the mini-jet contribution,  is given by the Fourier transform of overall soft gluon re-summation, i.e. we have
\begin{equation}
A_{BN}^{pp}(p;b,s)=\frac{e^{-h(p;b,s)}}{\int d^2 \textbf{b} e^{-h(p;b,s)} }
\end{equation}
where
\begin{equation}
h(p;b,s)=(const)\int_0^{q_{max}}\frac{dk_t}{k_t}
 \alpha_s(k_t) \log\frac{2q_{max}}{k_t}[1-J_0(bk_t)] \label{eq:hdb}
\end{equation}
 and
\begin{equation}
\alpha_s(k_t)\simeq (\frac{k_t}{\Lambda_{QCD}})^{-2p} \ \ \ \ \ \ k_t\rightarrow 0 \label{eq:alphabn}
\end{equation}
We have discussed the distribution of Eq.~(\ref{eq:hdb}) in many publications, its main characteristic is to 
include soft gluon re-summation down to $k_t=0$, and regulate the infrared singularity 
so as correspond to a dressed gluon potential $V(r) \sim r^{2p-1}$ for $r\rightarrow \infty$. The expression in Eq.(\ref{eq:alphabn}) is an ansatz put forward in \cite{Corsetti:1996wg} inspired by the Richardson potential behaviour at large distances \cite{Richardson}, and by Polyakov's argument \cite{Polyakov} about  linear Regge trajectories.

We have also shown an important consequence of  an expression such as the  
above for $\alpha_s(k_t\rightarrow 0)$ \cite{Grau:2009qx}, namely  that asymptotically  
the regularized and integrated soft gluon spectrum  of Eq.~(\ref{eq:hdb}) is seen to rise as
\begin{equation}
h(p;b,s)\rightarrow (b{\bar {\Lambda}})^{2p}
\end{equation}
and thus the $b-$ distribution exhibits
a cut-off in $b-$space strongly dependent on the parameter $p$, i.e. 
\begin{equation}
A_{BN}\rightarrow e^{-(b{\bar \Lambda})^{2p}} \ \ \ \  b\rightarrow \infty \label{eq:abnasymt}
\end{equation}
with ${\bar \Lambda}\propto \Lambda_{QCD}$. Since the mini jet cross-sections at low-x 
are parametrized so as to rise as $s^\epsilon$,  the behavior of Eq.~(\ref{eq:abnasymt}) 
leads to a high energy behavior for the  total cross-section given as 
\begin{equation}
\sigma_{tot}^{pp}\sim 
\frac{2 \pi}
{(\bar \Lambda)^2}[\epsilon \log s]^{1/p}
\end{equation}
The parameter  $1/2<p<1$: the lower limit so as to have  a confining potential, the upper limit 
to insure   convergence of the  integral over the soft gluon spectrum of Eq.~(\ref{eq:hdb}). An 
immediate consequence of this model is that the cross-section will never rise faster than $[\log s]^2$, 
the saturation of the Froissart limiting behavior being obtained for $p=1/2$. Notice, that, in this model, the  
mini-jet contribution, just as in hard Pomeron models \cite{Ryskin:2011qe},  rises as 
$\sigma_{jet}\sim s^\epsilon$, with $\epsilon\sim 0.3-0.4$ depending on the low-x parametrization 
of the PDF. However the strong cut-off in b-space
brought in by the singular, but integrable, 
effective $quark-soft-gluon$ coupling constant 
leads only up to a $(logarithmic)^2$ rise with energy.
For more details, we refer the reader to \cite{Grau:2009qx}. 

\begin{figure}
\centering
\resizebox{0.5\textwidth}{!}{
\includegraphics{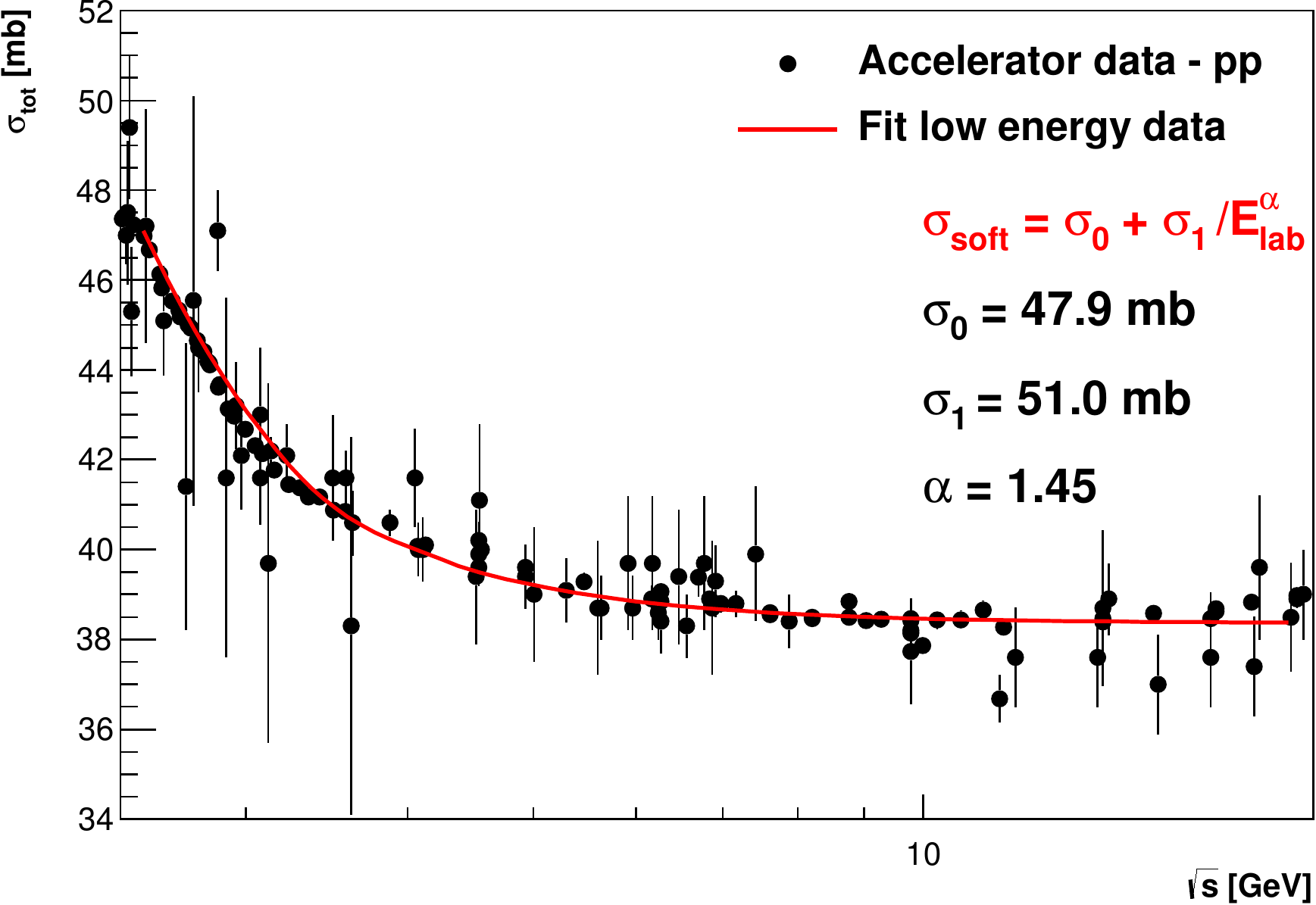}}
\caption{Low energy parametrization of $pp$ total cross-section}
\label{fig-pplow}
\end{figure}

The low energy term  includes collisions with $p_t \le p_{tmin}\sim (1\div 2)\ GeV$, and the cross-section 
$\sigma_{soft}^{pp}(s)$ is not predicted by this model so far, thus we parametrize it here with a 
constant and one or more decreasing terms. The result is shown in Fig.~ \ref{fig-pplow}. 

\begin{figure}
\centering
\resizebox{0.4\textwidth}{!}{
\includegraphics{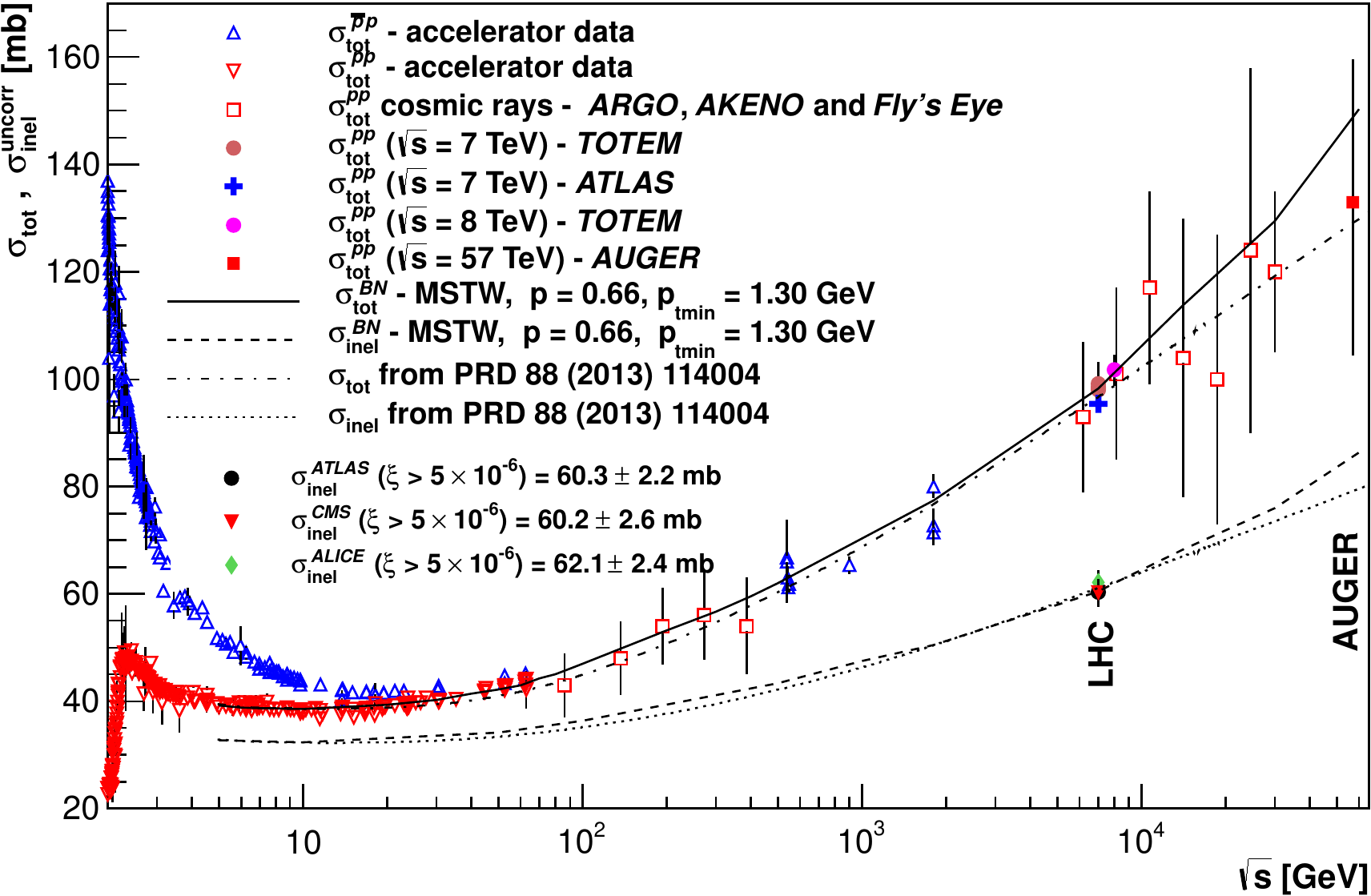}}
\caption{QCD mini-jet with soft gluon resummation model and $pp$ total cross-section (full line) 
as described in the text. The determination of the optimal model parameters was done in- dependently of the AUGER data. The dotted curve is from the mini-jet model of \cite{Giannini:2013jla}. 
 Accelerator data at LHC include TOTEM \cite{Antchev:2013paa,Antchev:2013iaa} and 
ATLAS  
\cite{Aad;2014dca}
 measurements.
The inelastic uncorrelated cross-section  
is given by the dashed curve and compared with central collisions results at LHC by 
ATLAS \cite{Aad:2011eu}, CMS \cite{Chatrchyan:2012nj}  and ALICE \cite{Abelev:2012sea}. }
\label{fig-ppall}
\end{figure}

The perturbative, mini-jet, part is  defined with $p_t^{parton}\ge p_{tmin}$ and is determined 
through a set of perturbative parameters for the jet cross-section, namely a choice of  PDF  
and $p_{tmin}$. Since the soft gluon re-summation includes all order terms in soft gluon emission, 
as in previous publications we have used only LO densities. An important point of our approach 
is that we use the same,  library distributed PDF, as used for jet physics.  Previously used PDFs 
were  GRV \cite{Gluck:1991ng,Gluck:1994uf,Gluck:1998xa}, or MRST72 \cite{Martin:1998sq}. 
Both  still give a good description of data up to LHC results, as shown here and in the next section.  
In Fig. \ref{fig-ppall} we  show the results  obtained through  a more    recent  set of LO densities, 
MSTW \cite{Martin:2009iq},  for both the total and the inelastic $pp$ cross-sections. 
 We notice here that the determination of the parameters, at low and high energy as well,  was done without  taking  account the cosmic ray data points, namely the parameters were chosen   so as to give good  reproduction of ISR and LHC data only. 
The parameter $p$, whose value is explicitly given  in this figure,  is related to the 
amount of 
 acollinearity induced by
   soft gluon emission, as discussed in \cite{Grau:2009qx}. 
Its value lies in the range $0.6\lesssim p \lesssim 0.8$ depending on the PDF used. 
For MSTW, we find that the parameter set  \{$p_{tmin}=1.3\ GeV$, p= 0.66\}  best 
reproduces the $pp$ cross-section up to LHC8.

We note the important result that the inelastic cross-section predicted by the  
parametrization of the total cross-section through a single-channel eikonal, reproduces very well the LHC 
data for non-diffractive collisions by ATLAS \cite{Aad:2011eu}, CMS \cite{Chatrchyan:2012nj}  
and ALICE \cite{Abelev:2012sea}. Such agreement had already been highlighted in  \cite{Achilli:2011sw}. 
We shall return to comment on this point at the end of the paper. 

 In addition to the results from our model, Fig.~\ref{fig-ppall}  shows also the mini-jet result Ref. ~\cite{Giannini:2013jla}, where a different set of PDFs is used, and a different impact parameter distribution. The mini-jet contributions in these two applications of the mini-jet model are different, but both are based on a  single-channel eikonal approach.
\subsection{A comment on the model parameters}
The present focus of our model 
is the parametrization of the   high energy behavior described by QCD processes.  
To this aim, we need a set of PDFs, a lower cut-off dividing the perturbative and non-perturbative 
regions, $p_{tmin}$, and a 
parameter $p$, which we also referred to as {\it singularity} parameter. 
The higher this parameter, the more 
softened is the cross-section. 
Phenomenologically, its value  
is fixed in relation  to the low-x behavior of the densities.
The parameter $p$ thus appears to be unrelated to the perturbative expression for the  
QCD coupling constant $\alpha_s(Q^2)$. We however believe it to be of more fundamental interest, 
and have made the ansatz \cite{Pancheri:2014rga} that  the actual 
expression to use in the integrand of Eq. ~(\ref{eq:hdb}) is
 \begin{equation}
 \label{eq:BN}
 \alpha_s^{BN}(Q^2)=\frac{1}{\ln[1+(\frac{Q^2}{\Lambda^2})^{b_0}]} \overset{Q^2>>\Lambda^2} {\longrightarrow}\alpha_{AF}(Q^2)
 \end{equation}
 where $b_0=(33-2N_f)/12 \pi$ and the  suffix {\it BN} is used to indicate its applicability  
 into the infrared region (the one first explored in QED  by Bloch and Nordsieck \cite{Bloch:1937pw}), 
 while coinciding with the usual one-loop asymptotic freedom expression at high $Q^2$.  
 The above ansatz  would imply that the infrared region description  does not require introduction 
 of an extra parameter $p$:  the behavior from $Q^2=0$ to $Q^2\to \infty$ is dictated only by  the anomalous dimension factor. 
However, the present uncertainty about a fundamental calculation for the low-x behavior 
of the parton densities, prevents a full use of Eq.(\ref{eq:BN}). Suffice to say that  our 
phenomenological values for $p$ are in the same range of variability of the anomalous dimension factor $b_0$.
\section{The production cross-section for $p-air$}
\begin{figure}
\centering
\resizebox{0.5\textwidth}{!}{
\includegraphics{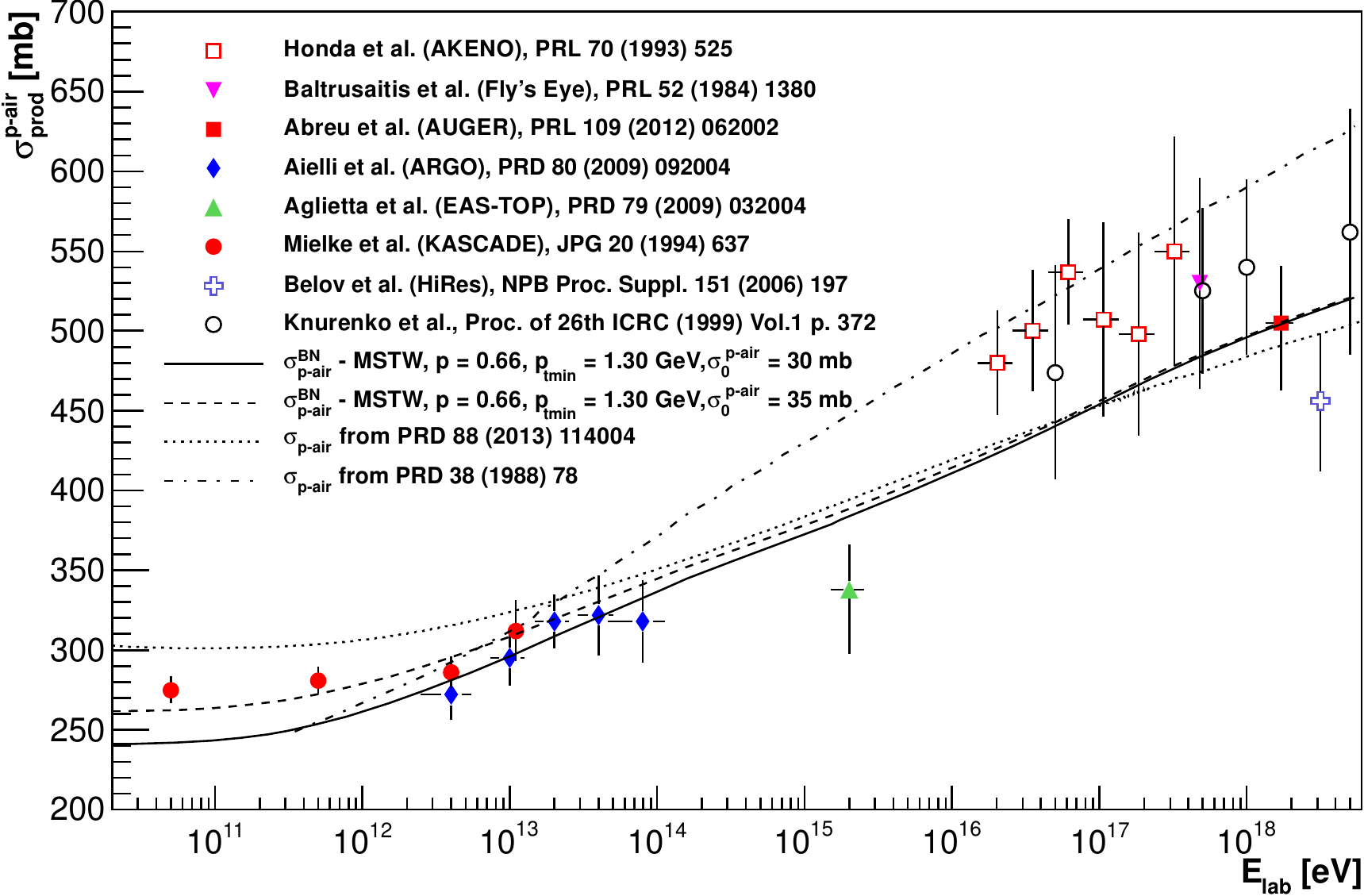}}
\caption{ $p-air$ production cross-section using MSTW2008 \cite{Martin:2009iq} parton densities in a single-channel eikonal 
mini-jet model with infrared gluon resummation. 
The two curves from the BN model (full and dashes) are obtained with a different low energy constant, $\sigma_0$. The results from the mini-jet model of \cite{Giannini:2013jla} is given by the dotted curve. The dashed curve is the first mini-jet application of a mini-jet model for  cosmic ray cross-section, from \cite{Durand:1988ax}.}
\label{fig-cosmic1}
\end{figure}

With  the low energy part parametrized as shown in Fig.~\ref{fig-pplow}, and the mini-jet part, 
we can calculate the inelastic $pp$ cross-section and thus the production $p-air$ cross-section.  
The result is shown in Fig. ~\ref{fig-cosmic1} where our model is compared with cosmic ray data 
\cite{Collaboration:2012wt,Baltrusaitis:1984ka,Honda:1992kv,Mielke:1994un,Knurenko:1999cr,
Belov:2006mb,Aglietta:2009zza,Aielli:2009ca} 
and with two other mini-jet models, a recent one  \cite{Giannini:2013jla} and the first such application by Durand and Pi \cite{Durand:1988ax}. We notice that the old model is rather above the cosmic ray data point, as they are presently extracted. 
In this figure we have reduced the  constants $\sigma_{0,1}$ in the $pp$ cross-section so as to comply 
with the surface/volume effect for the low transverse momentum collisions. Because of the uncertainty 
in this low energy region, the soft term in the $pp$ cross-section has been included openly as a constant.  
However, we have also considered the full low-energy parametrization of Figs. \ref{fig-pplow},\ref{fig-ppall}, but
in the energy range of Fig. \ref{fig-cosmic1}  such low energy  decreasing term makes no difference whatsoever.

\begin{figure*}
\resizebox{1.0\textwidth}{!}{
\includegraphics{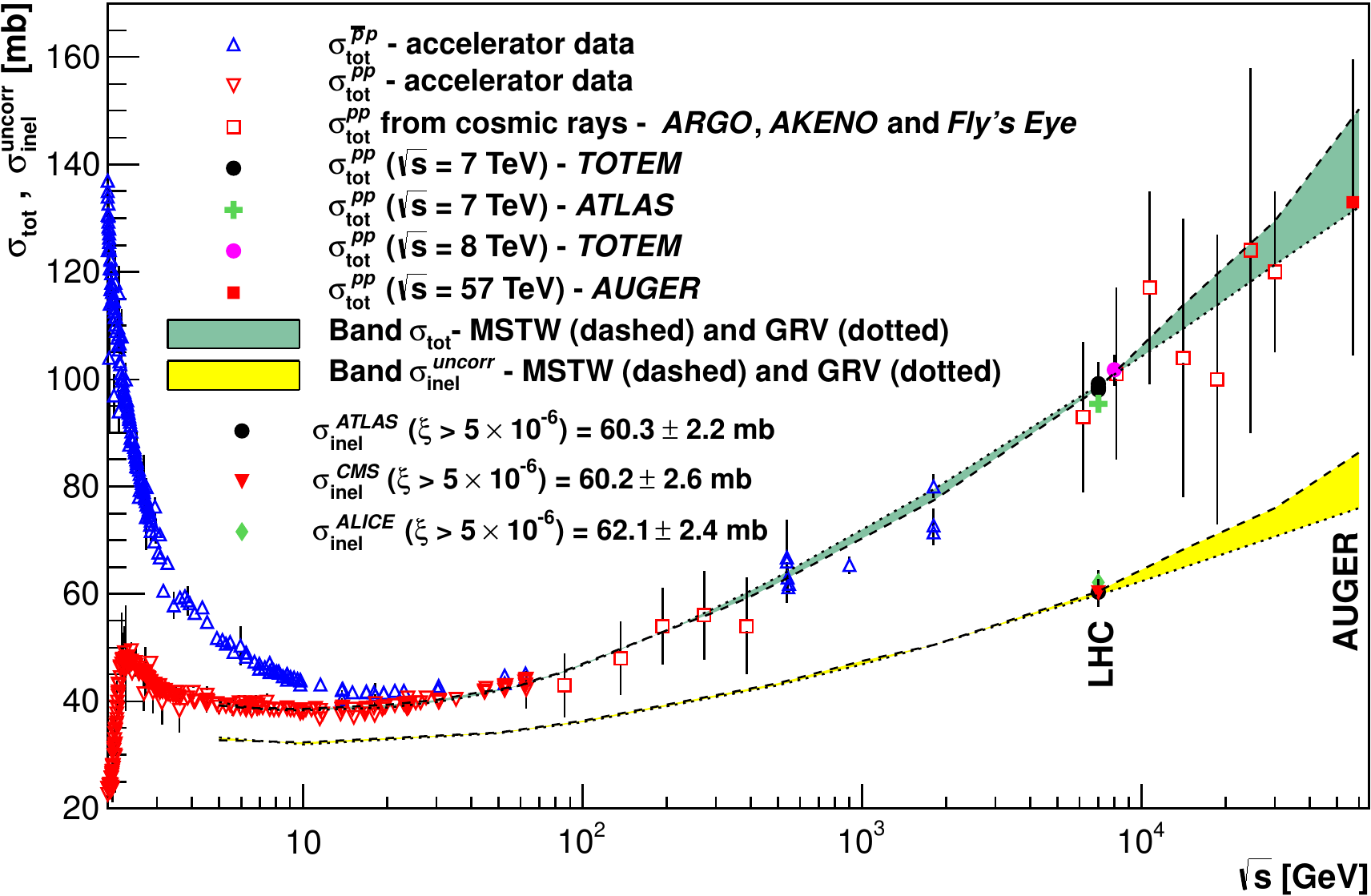}
\includegraphics{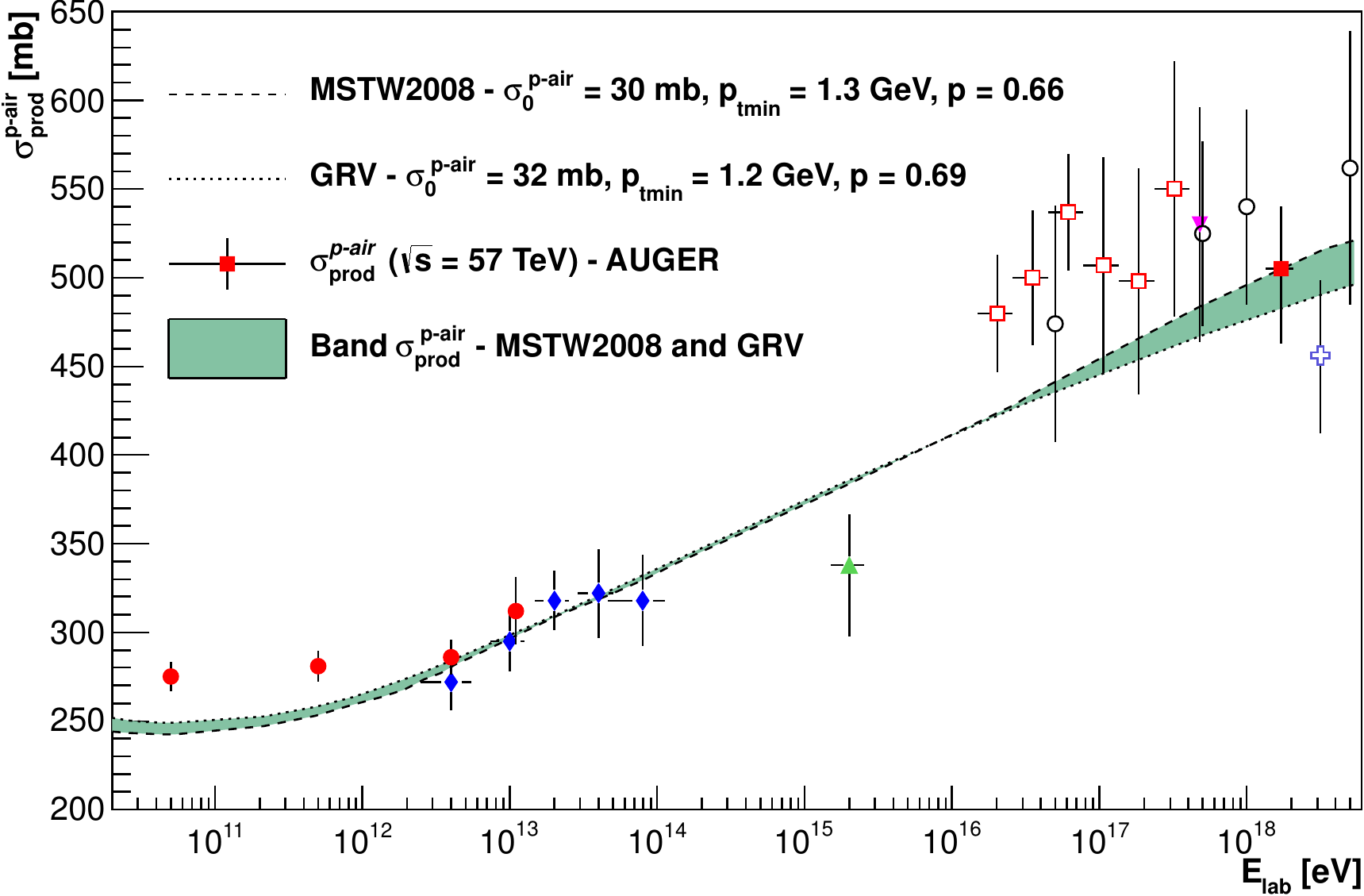}}
\vspace*{1cm}
\caption{Left panel: total and uncorrelated  inelastic  $pp$ cross-sections for different PDF sets.  
Right panel: the production $p-air$ cross-section following the results from the left panel. 
The green and yellow bands indicate the uncertainty related to the low-x behavior of the PDF used. 
Symbols for $p-air$ data  as in Fig. \ref{fig-cosmic1}.}
\label{fig:cosmic2}
\end{figure*}

To estimate  the error of this procedure as well as check the stability of the model   and its application to 
both  $pp$ and $p-air$ cross-sections, we have done the following checks:
\begin{itemize} 
\item  after parametrizing   the low energy part of  $pp$ data,  
the rise has been described through other  available LO PDFs, namely    MRST72 and GRV in addition to 
MSTW. For a given PDF set, the
parameters $p_{tmin}$ and $p$ have been chosen to best  reproduce  LHC results for 
$\sigma^{pp}_{tot}$  \cite{Antchev:2013paa,Antchev:2013iaa} .
\item we have done an actual fit to both the low energy data and LHC (excluding cosmic rays extracted data), 
using GRV and MRST72, and with the free 
singularity
 parameter $p$.
\item  We have changed the nuclear density model, applying a Wood-Saxon potential, also applied,  for instance, in   \cite{Gotsman:2013nya} and \cite{Giannini:2013jla}.
\end{itemize}
The results of this  exercise for different densities are     shown in the two panels of Fig.
  \ref{fig:cosmic2},  where the bands highlight the uncertainty related to the the low-x behavior of the 
  parton densities used for the mini jet calculation. 
As we are not so much interested in understanding right now the low energy part, the constant 
$\sigma_0$ has simply been reduced adjusting it to the data.  As expected, the contribution 
from the low energy part gets weaker and weaker for very high energies.
The results are also  shown in Table \ref{tab:xsections}.
\begin{table}
\caption{Total and inelastic uncorrelated  $pp $ cross-sections (second and third column). 
Fourth column is   the uncorrelated inelastic $pp$ cross-section  for input to  the Glauber formula 
for $\sigma^{p-air}$, with low energy part reduced for nuclear area/volume effect.  
Last column shows  the resulting $p-air$  cross-section. Different parameter sets are as indicated.} \label{tab:xsections}
 \tiny
\centering
\begin{tabular}{|c|c|c|c|c|} \hline
\multicolumn{5}{|c|}
{Parameter set : GRV, $p_{tmin}=1.2\ GeV$, p=0.69}\\ \hline
         $\sqrt{s} \ (GeV)$ & $\sigma_{tot}^{pp}$ &  $\sigma^{pp-uncorr}_{inel}$     &$\sigma^{pp-uncorr}_{inel}$        &$\sigma^{p-air}_{prod}$\\ 
                                      & with $\sigma_0=48\ mb$& with $\sigma_0=48\ mb$ &with $\sigma_0= 32\ mb$&$\sigma_0= 32\ mb$ \\
         \hline
 5 & 		39.9	& 33.2	& 24.9	&  255.8\\
 10 &		38.2	& 32.0	& 24.0	&248.9\\
50 & 		41.9	& 34.0	&26.7 	&268.7\\
100 & 	46.7&36.1	&29.7	&288.6\\
500& 	63.2	& 43.0	& 38.6	&340.9\\
1000 . & 	71.7	& 46.9	& 43.1 	&364.1\\
1800. & 	79.5	& 50.5	& 47.2	&383.5\\
7000 &	98.9 	& 59.8	& 57.4	&426.1\\
8000 & 100.9	& 60.7	&58.4		&430.0\\
14000 & 109.3	&  64.8	&62.8		&445.9\\
30000 & 121.3	& 70.7	&69.0		&467.0\\
60000 & 132.0	& 76.0	&74.6		&484.3\\
 \hline \hline
 \multicolumn{5}{|c|}{
 Parameter set : MRST72, $p_{tmin}=1.25\ GeV$, p=0.62}\\
  \hline
 5   &		39.9		&33.2	& 24.9	&   255.8\\
 10 &		38.3		& 32.0	& 24.0	&  249.1\\
 50 &		43.1	& 34.6	& 	27.6		& 274.5\\
 100 &	48.4  &	36.9 & 	30.8		&  295.8\\
 500&	63.8  &43.7	&	39.3		& 344.6\\ 
 1000 &	71.3  &	47.1&	43.3		& 365.1\\
 1800 &	78.1  &	50.3 &46.9		& 382.3\\
 7000 &	98.2  & 60.4	&58.0	&  428.3\\
 8000 &	100.7 &61.7	&59.4	& 433.6\\
 14000 &	112.2&67.7	&65.7		& 456.2\\
 30000 &	129.1&76.5	&75.0		& 485.7\\
 60000 &	144.2&84.4	&83.3		& 509.1\\
\hline \hline
 \multicolumn{5}{|c|}{ Parameter set : MSTW, $p_{tmin}=1.3\ GeV$, p=0.66}\\ 
 & with $\sigma_0=47.9 \ mb$& with $\sigma_0=47.9\ mb$&with $\sigma_0= 30\  mb$&$\sigma_0= 30\ mb$.\\
 \hline
5 &	39.21 	& 32.7 	&23.7 &246.8 \\
10 &	38.60 	& 32.3 &23.1	& 242.6 \\
50 & 		42.2 	& 34.2 & 	25.9	&  263.4\\
100 &46.9 	&  36.4& 29.2&285.5 \\
500&	62.0& 43.3	&38.1 	&338.6\\
1000&	71.0 & 47.5	&43.1	&364.4\\
1800	&	77.5 	& 50.5	&46.6&381.2 \\
7000	 & 98.3	& 60.5	& 57.8	& 428.0  \\
8000 & 101.3	&	62.0	&59.4	     &434.0\\
14000 & 113.7 & 68.2	&66.1	     &457.7	   \\
30000 & 129.4 & 76.0	&	  74.3   & 483.8	   \\
60000 & 150.3 & 86.3	&     85.1       & 514.3	   \\
 \hline \hline

 \end{tabular}
\end{table}
In the table,  the low energy part of the eikonal function,  $n_{soft}$,   is fitted to the 
low energy data alone, as in Fig.~\ref{fig-pplow}, whereas the QCD part $n_{hard}$ is 
chosen so as to best describe  the $pp$ accelerator data. As for the  other  check, 
non reproduced in this table, namely fitting   at the same time both the low and the 
high  energy accelerator data in order to determine the best $p$-value, for a given  
choice of PDF and $p_{tmin}$, we have found the result to be  consistent with above, 
for  $p\simeq 0.6$ for MRST72  densities and $p_{tmin} \simeq (1.3\div 1.4)\ GeV$.  
Using the Wood-Saxon potential slightly lowers the curves for $p-air$ with respect to the standard 
nuclear potential of Eq. (\ref{eq:nucl_prof}).
Before concluding, we would like to return to an important physics point,
 namely 
that the experimentally measured $\sigma^{prod}_{p-air}$ differs from the total 
$\sigma^{tot}_{p-air}$ through the exclusion of elastic $\sigma^{el}_{p-air}$ as well 
as quasi-elastic $\sigma^{q-el}_{p-air}$. An example of $\sigma^{q-el}_{p-air}$ 
is given by processes such as $p + N \to\ p^* + N$. 

  In general, one has to carefully examine the contribution that  the measured  cosmic ray particle production cross-section receives from  (single as well as double) diffractive processes. It is possible that at the very high energies reached by present day cosmic ray experiments, and as  proposed in \cite{Block:2006hy}, diffraction needs not to be included.  
This acquires a particular significance (and endows a certain simplicity) to (single-channel) 
mini-jet models when applied to an analysis of cosmic ray cross-sections. As we have discussed, 
the inelastic cross-section in such models only includes uncorrelated process, and  does   not include the diffractive part.  
Our present proposal is that 
the single-channel approach  is thus  best suited for calculations of  the production $p-air$ cross-section from  cosmic ray measurements at ultra high energies.
For this purpose, we have employed parameters (such as $p$) suitable for describing the 
total cross-section well and by default giving us the inelastic part devoid of diffraction. 
A posteriori, such a description seems to work quite well.

The most remarkable result that we find is that we reproduce very well the AUGER point, 
in addition to have a reasonably good description of all the more recent cosmic ray measurements. 
At the AUGER point, our result agrees with predictions of recent (much more complex) MC interaction codes, such as QGSJET01c \cite{Knurenko:2011aa}. We notice once more that our result is obtained in the context of a single-channel eikonal formalism and a good description of the accelerator data for pp total cross- section,  to which we arrived  without attempting to reproduce the extracted $pp$ value at cosmic ray data energies.

\section{Concluding remarks}
In this contribution, we have seen
that the Glauber formula in conjunction with an inelastic $pp$ cross-section obtained 
through a single-channel eikonal formalism provides a very good description of the 
cosmic ray extracted ($p-air$) cross-section. Thus, we might ask, whether a single-channel 
eikonal expression adequately representing the $pp$ total cross-section is also sufficient 
to describe high energy elastic scattering. Obviously not.
However, it is fair to say that the momentum transfer ($t$)-dependence of the elastic differential 
cross-section from the forward ($t =0$) up to after the dip still escapes a fundamental 
QCD explanation. For this, and thus for the diffractive part of the cross-section, 
a multi channel formalism  \cite{Khoze:2000wk,Lipari:2009rm,Gotsman:2012rq}
is still required. However, it is our ansatz that a viable multichannel formalism must be 
geared to reproduce the results from a single term at the optical point (that is at $t =0$).

For the present, we may reiterate that a good 
single-channel eikonal representation for the 
total cross-section should be sufficient to describe the cosmic ray $p-air$     
production cross-section data and conversely, that models which reproduce $\sigma^{p-air}_{production}$ 
can be trusted to extrapolate correctly $\sigma^{pp}_{inel-non-diffractive}$, and thus the total 
$\sigma^{pp}_{tot}$ in a single-channel eikonal model. However,  very high energy predictions are 
affected by an uncertainty related to the low-x behavior of the PDFs used in the phenomenological 
calculation of the mini-jet cross-sections. It may thus be very important to include the forthcoming 
LHC data at $\sqrt{s}>10\ TeV$ to reduce such uncertainty and hopefully be  able to extract information 
on $\sigma^{pp}_{tot/inel}$ from the even higher energy cosmic ray measurements to be expected from cosmic rays.

 \section*{Acknowledgments}
A.G. acknowledges partial support by Junta de Andalucia (FQM 6552, FQM 101).
 D.A.F. acknowledges the S{\~a}o Paulo Research Foundation (FAPESP)  
and the Coordination for the Improvement of Higher Education Personnel (CAPES) for financial support (contract: 2014/00337-8).  O.S. acknowledges partial support from funds of Foundation of Polish Science
grant POMOST/2013-7/12, that is co-financed from European Union, Regional
Development
Fund.




\end{document}